\documentclass[sigconf]{aamas}  

\usepackage{booktabs}
\usepackage{comment}
\usepackage[ruled,vlined,linesnumbered,noresetcount]{algorithm2e}
\usepackage{xcolor,colortbl}

\setcopyright{ifaamas}  
\acmDOI{doi}  
\acmISBN{}  
\acmConference[AAMAS'19]{Proc.\@ of the 18th International Conference on Autonomous Agents and Multiagent Systems (AAMAS 2019), N.~Agmon, M.~E.~Taylor, E.~Elkind, M.~Veloso (eds.)}{May 2019}{Montreal, Canada}  
\acmYear{2019}  
\copyrightyear{2019}  
\acmPrice{}  

\usepackage{bbm}
\usepackage[bbgreekl]{mathbbol}
\usepackage{amsfonts}
\usepackage{amsmath,bm}
\usepackage{amsthm,amssymb}
\everymath{\displaystyle}
\usepackage{subfigure}

\newtheoremstyle{thm-sf}{}{}{\itshape}{}{\sffamily\bfseries}{.}{ }{}
\theoremstyle{thm-sf}

\newcommand{\RNum}[1]{\uppercase\expandafter{\romannumeral #1\relax}}
\newcommand{\sets}{\mathbb}
\newcommand{\random}{\mathcal}

\usepackage{enumitem}   

\begin{document}

\title{Social Network Based Substance Abuse Prevention via Network Modification (A Preliminary Study)} 




%
\author{Aida Rahmattalabi}
\affiliation{%
\institution{University of Southern California}
}
\email{rahmatta@usc.edu}

\author{Anamika Barman Adhikari}
\affiliation{%
\institution{University of Denver}
}
\email{anamika.barmanadhikari@du.edu}

\author{Phebe Vayanos}
\affiliation{%
\institution{University of Southern California}
}
\email{phebe.vayanos@usc.edu}

\author{Milind Tambe}
\affiliation{%
\institution{University of Southern California}
}
\email{tambe@usc.edu}

\author{Eric Rice}
\affiliation{%
\institution{University of Southern California}
}
\email{ericr@usc.edu@usc.edu}

\author{Robin Baker}
\affiliation{%
\institution{Urban Peak Organization}
}
\email{Robin.Baker@urbanpeak.org}

%
%
%
%
%
%
%
\begin{abstract} 
Substance use and abuse is a significant public health problem in the United States. Group-based intervention programs offer a promising means of preventing and reducing substance abuse. While effective, unfortunately, inappropriate intervention groups can result in an increase in deviant behaviors among participants, a process known as \emph{deviancy training}. 
This paper investigates the problem of optimizing the social influence related to the deviant behavior via careful construction of the intervention groups.
We propose a Mixed Integer Optimization formulation that decides on the intervention groups to be formed, captures the impact of the intervention groups on the structure of the social network, and models the impact of these changes on behavior propagation. In addition, we propose a scalable hybrid meta-heuristic algorithm that combines Mixed Integer Programming and Large Neighborhood Search to find near-optimal network partitions.
Our algorithm is packaged in the form of GUIDE, an AI-based decision aid that recommends intervention groups. Being the first quantitative decision aid of this kind, GUIDE is able to assist practitioners, in particular social workers, in three key areas: \emph{(a)} GUIDE proposes near-optimal solutions that are shown, via extensive simulations, to significantly improve over the traditional qualitative practices for forming intervention groups; \emph{(b)} GUIDE is able to identify circumstances when an intervention will lead to deviancy training, thus saving time, money, and effort; \emph{(c)} GUIDE can evaluate current strategies of group formation and discard strategies that will lead to deviancy training. In developing GUIDE, we are primarily interested in substance use interventions among homeless youth as a high risk and vulnerable population. GUIDE is developed in collaboration with Urban Peak, a homeless-youth serving organization in Denver, CO, and is under preparation for deployment. 
\end{abstract}

\keywords{Social Networks; Substance Abuse Intervention, Optimization}  
\maketitle
\section{Introduction}

Peers have a direct influence in adolescents' risk behaviours
Substance use and abuse is a significant public health problem in the United States, particularly among youth. According to the Monitoring the Future study \cite{mccabe2014social}, around 54 percent of high school students have tried at least one illicit substance. Homeless youth, in particular, are shown to be disproportionately affected, with substantially higher levels of substance use compared to the housed youth \cite{nyamathi2012characteristics}. 

\begin{figure}
 \includegraphics[width=0.47\textwidth]{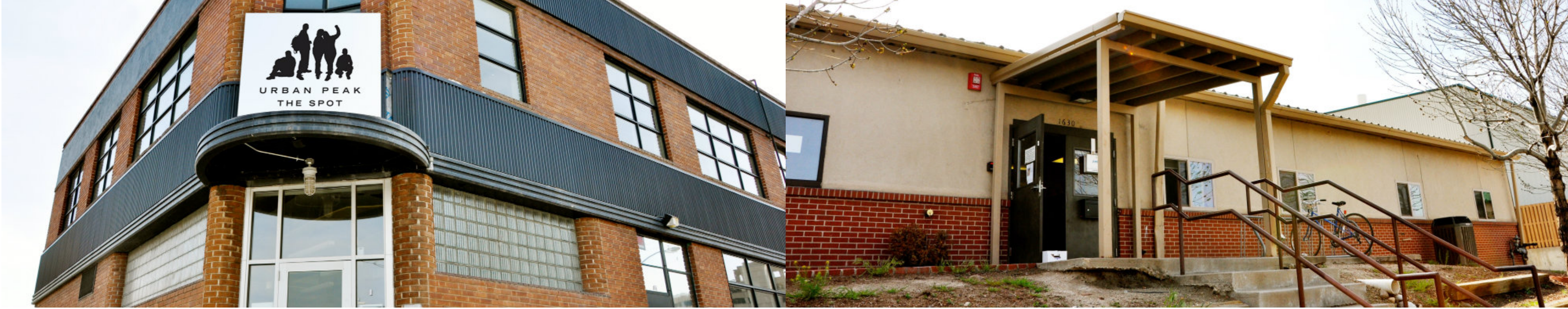}
 \caption{Urban Peak site. We are collaborating with Urban Peak organization for conducting the group-based interventions}
 \label{fig: shelter}
\end{figure}

Notably, interventions attempting to reduce substance use have successfully utilized social networks to disseminate and reinforce behavioral norms supportive of protective behaviors related to substance use (e.g., \cite{valente2003effects}). The way the social network is utilized is often through formation of subgroups where the individuals can talk, share experiences and engage in various constructive activities. Such social network-based approaches to substance abuse prevention are considered more promising because young people learn better from one another, find each other more credible and have a shared environment and culture \cite{georgie2016peer}. 

Unfortunately, these social network-based efforts may inadvertently increase the chances of youth being exposed to negative social influences, as they do not explicitly structure the intervention groups. This can result in an effect known as deviancy training. Deviancy training occurs
when high-risk youth are aggregated together and reinforce negative behaviors and attitudes. 

Social network-based interventions have typically grouped participants into intervention groups in three ways~\cite{valente2003effects}: (1) \textit{random assignment}: participants are randomly assigned to the groups; 
(2) \textit{network based assignment}: participants are assigned based on their own nominations;~(3) \textit{teacher nominated assignment}: groups are created based on teacher nominations, often corresponding to an assignment based on the participants' behaviors, i.e., even distribution of high-risk individuals. The random assignment and teacher nominated methods are less effective because they are not designed to leverage naturally occurring ties in participants' social networks. The network-based method on the other hand does leverage these natural forms of influence. However, it is shown that it can have unintended effect of exacerbating the problem behavior~\cite{valente2003effects}. 

Recently, successful applications of AI based decision aids such as~\cite{yadav2016using} and~\cite{dickerson2012optimizing} have encouraged efforts to address such complicated social problems using techniques in AI and optimization~\cite{, 10.1007/978-3-030-01554-1_35}. 
In the present work, we aim to tackle the problem of deviancy training in substance abuse intervention by structuring more effective groups, one that effectively partitions the participants' social network.

Building this decision aid, however, raises major challenges. 

First, it has been observed that in the process of being a part of the network-based intervention, the sub-networks (peer-groups) undergo a transformation, whereby network ties change in strength~\cite{spoor2004evolutionary}. Such changes will consequently change the influence spread patterns and therefore constitute an important component of the model.
Second challenge is understanding the influence process. In this regard, we propose an influence spread model which is based on the competitive Linear Threshold (LT) Model in~\cite{borodin2010threshold} to explain both positive and negative social influences within interventions. The third major challenge is to strategically choose the right peer-groups from the larger social network to mitigate the challenge of deviancy training. In the following, we will refer to those engaged in substance use as ``users'', that we wish to have socially influenced by 
``non-users'' peers in the smaller groups, so as to reduce/prevent such substance use and abuse.

We view this problem as an influence maximization problem, with the goal of maximizing/minimizing the total positive/negative influence by careful construction of the intervention groups, considering the consequent changes in the network topology. 
Influence maximization in networks has been widely studied in the past decade and several works have tried to address the problem of modifying the network topology in order to maximize positive influence spread or curtail undesirable behaviors. In particular, \cite{kimura2008solving,KhaDilSon14} study the problem of edge deletion/addition under the LT model. Our problem however is different: we do not directly decide on edges to add/delete. Instead, we can only select how participants are partitioned into intervention groups. Depending on this choice, there is a disciplined process that controls how social network changes. More precisely, we will cluster the given social network into intervention groups, which causes further changes in the network topology as a byproduct, and subsequently controls how the positive and negative influences propagate.


Social influence optimization and network dynamics distinguish this work from the existing literature on network clustering. For example, ~\cite{bansal2004correlation} studies clustering graphs with both positive and negative edges, but the structure of the network is static. Further, unlike ~\cite{gorke2009dynamic} that considers the network dynamic, we are interested in optimizing a stochastic influence function, plus, the evolution of the network topology is not just a function of time but a function of the decisions variables or clusters. To best of our knowledge no work has addressed such influence-based clustering of networks to this date. 

To address these challenges, we propose an AI-based decision aid, called GUIDE (\textbf{G}ro\textbf{U}p-based \textbf{I}ntervention \textbf{DE}cision aid). GUIDE assists interventionists in substance abuse prevention by giving recommendations regarding the intervention groups. In what follows we detail our main contributions: (\RNum{1}) First, we introduce a novel problem to AI researchers; We propose a mathematical model for this type of intervention, which includes the key aspects of the influence spread and network dynamics. The model enables us to predict, both the expected success of the intervention, measured as the expected number of ``non-users'' at the end of the intervention, and the possibility of harm, or deviancy training. 
(\RNum{2}) We show that the problem of finding the optimal partition to minimize the expected substance use is NP-hard. (\RNum{3}) Therefore, we propose both a Mixed Integer Linear Program (MILP) to solve this problem and a hybrid meta-heuristic algorithm that combines MIP and Large Neighborhood Search (LNS) to find near-optimal network partitions. (\RNum{4}) We provide extensive analysis of the model, using both real-world social network data and synthetic graphs and we show that our proposed partitions can significantly outperform all common practices such as random assignment of individuals or participants choice.

\section{Problem Statement}
Our goal is to create an assistant that will aid an interventionist for substance abuse prevention. The intervention that we consider is a 6-week program, in which participants are placed into different groups. The size of each group is bounded between 4 and 7. An interventionist interacts with each group. 
Pre-intervention, the participants are asked to report both their hard-drug using behaviors, and their network relationships (i.e., who they know) and what the strength of that relationship is (whether a person is a strong or a weak tie). Over the course of the intervention and based on the groups assignments, these social ties will change: some may become weaker, while others may grow stronger. We aim to maximize the number of 
``non-users'' one-month after the intervention is complete.

We view the problem of group configuration for substance abuse prevention as a graph
partitioning problem. Given a social network, the goal is to find a partition such that the expected number of ``non-users,'' at the end of the intervention, is maximized. More formally, let graph $\sets G = (\sets V,\sets E)$ be a directed graph representing the given social network, with~$\sets V$ as the set of all nodes (individuals in the social network), and~$\sets E \subset \sets V \times \sets V$ as the set of all edges (social ties). Arc $(i,j) \in \sets E$ indicates existence of an arc pointing from $i$ to $j$, signifying that $j$ has reported $i$ as a friend pre-intervention. Associated with each node $i \in \sets V$ is a \emph{node behavior} indicating that node $i$ is a substance ``user'' or a ``non-user.'' The scalar $b_i \in \{0,1\}$ encodes the node behavior with $b_i=1$ if and only if node $i$ is a ``user.'' Additionally, associated with each edge $e=(i,j) \in \sets E$ is the \emph{edge strength} $s_e \in \{0, 1\}$, where $s_e = 1$ (resp.\ 0), indicates that~$j$ considers~$i$ as strongly (resp.\ weakly) influential for him.

We model the intervention as a partition of $\sets V$ into $S$ subsets~$\sets P_s$, $s=1,\ldots,S$, such that $\cap_{s=1}^S \sets P_s = \emptyset$, $\cup_{s=1}^S \sets P_s = \sets V$, where each subset must consist of at least $\underline C$ and at most $\overline C$ nodes. Note that the number of subsets $S$ is a decision variable, but $\underline C$ and $\overline C$ are pre-specified. 
During the intervention and based on the group configurations, the ties undergo changes. Some ties are cut or weakened and some will be reinforced. We assume that, given a choice of partition $\sets P := \{ \sets P_s \}_{s=1}^S$, the post-intervention structure of the network is known deterministically. We let $\sets G^+(\sets P) = (\sets V,\sets E^+(\sets P))$ denote the post-intervention graph when groups are formed according to partition $\sets P$, with $\sets E^+(\sets P)$ corresponding to the new edge set. Thus,
$(i, j) \in \sets E^+(\sets P)$ if and only if $j$ considers $i$ as a friend post-intervention. Accordingly, for $e\in \sets E^+(\sets P)$, we denote by $s_e^+ (\sets P) \in \{0,1\}$ the strength of edge~$e$ post-intervention. 

\begin{figure}
\centering
\includegraphics[width=1.0\linewidth]
{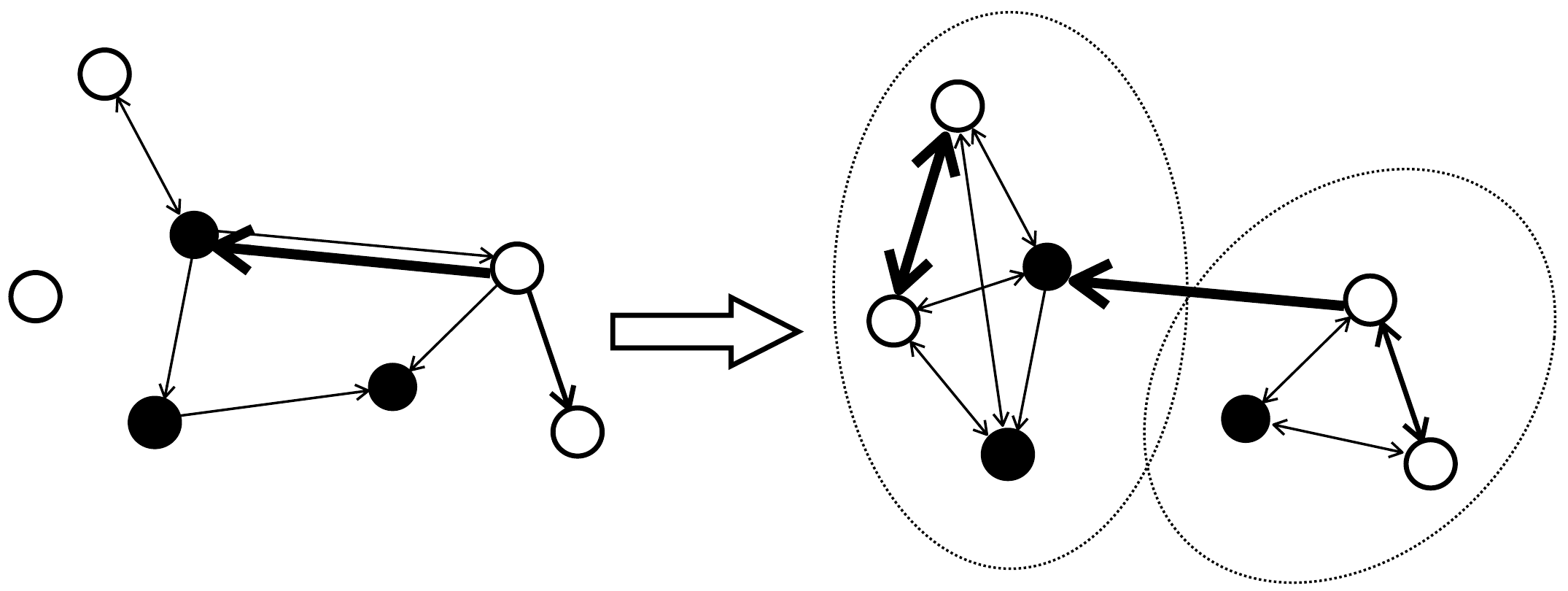}
\caption{An example network pre- (left) and post- (right) intervention. The black (resp. white) circles indicate ``user'' (resp. ``non-user'') nodes. Weak (resp. strong) links are denoted by thin (resp. thick) arrows. The ellipsoids represent the two groups that are formed for the intervention. As seen, new edges are created within the groups, while some edges are cut across the groups.  partitions}
\label{fig:network}
\end{figure}

The behavior of node $i\in \sets V$ post-intervention is random. We let $\random B_i(\sets P)$ denote the random variable that represents the behavior of node $i$ post-intervention, i.e., $\random B_i(\sets P)=1$ if and only if node $i$ is a ``user.'' This is a complicated stochastic function of the partition that depends on both the link formation and breakage rules and the influence model assumed. Figure \ref{fig:network} shows an example network pre- and post-intervention. \\
\textbf{Mathematical Formulation. } Mathematically, the problem of selecting the optimal partition $\sets P$ that maximizes the expected number of ``non-users'' in the network post-intervention can be formulated as:
\begin{equation}
\renewcommand{\arraystretch}{1.5}
\begin{array}{cl}
\text{maximize} & \displaystyle E\left[ \sum_{i \in \sets V} (1-\random B_i(\sets P)) \right] \\
\text{subject to} & S \in \mathbb N_+ \\
& \sets P_s \subset \sets V \quad \forall s \in \{1,\ldots,S\}\\
& \displaystyle \bigcap_{s=1}^S \sets P_s = \emptyset \\
& \displaystyle \bigcup_{s=1}^S \sets P_s = \sets V \\
& \displaystyle \underline C \leq | \sets P_s | \leq \overline C \quad \forall s \in \{1,\ldots,S\},
\end{array}
\label{eq:general_formulation}
\end{equation}
Where $E[\cdot]$ denotes the expectation operator with respect to the distribution of $\{\random B_i\}_{i \in \sets V}$. The objective counts the expected number of ``non-users'' in $\sets G^+(\sets P)$. Note that we are considering the intervention period as one time step and therefore, we are assuming a single-stage influence process.
The first three constraints ensure that $\sets P$ forms a partition of the node set. The last constraint enforces capacity constraints on each group. 

Problem~\eqref{eq:general_formulation} is a combinatorial optimization problem which is hard to solve as formalized by Theorem~\ref{thm:NP_hard}. In order to solve this problem, both the random behavior $\random B$, as well as the network structural changes must be modeled. In the following two sections, we propose models for network and influence dynamics that are supported by the social work literature.\\
\textbf{Tie Formation and Breakage. }
\label{par: net_dynamic}
\begin{table}
\begin{center}
\begin{tabular}{|c|c|c|c|}
\hline
\cellcolor[gray]{0.8}Same Group & \cellcolor[gray]{0.95}no-tie & \cellcolor[gray]{0.95}weak & \cellcolor[gray]{0.95}strong \\
\hline
(user, user) & strong & strong & strong \\
(non-user, non-user) & strong & strong & strong \\
(non-user, user) & weak & weak & strong \\
(user, non-user) & weak & weak & strong \\ \hline
\cellcolor[gray]{0.8}Separate Groups & \cellcolor[gray]{0.95}no-tie & \cellcolor[gray]{0.95}weak & \cellcolor[gray]{0.95}strong \\
\hline
(user, user) & none & none & strong \\
(non-user, non-user) & none & weak & strong \\
(non-user, user) & none & none & weak\\
(user, non-user) & none & none & weak \\
\hline
\end{tabular}
\end{center}
\caption{Changes in tie strength post-intervention. The existing relationships, and the behavior of the individuals as well as their assignment to groups impacts the changes.}\label{table:net_in_out}
\end{table}
While these youth have pre-existing strong or weak ties, especially centered around their substance use, the strength of these relationships would weaken or strengthen during the intervention \cite{centola2007complex}. 
There is empirical evidence to suggest that the more similar two individuals are, the stronger their ties are \cite{aral2014tie}. This is explained by their behavioral homophily. Homophily refers to the tendency of people to associate with people who are like themselves and these ties also often facilitate more communication and influence \cite{mcpherson2001birds}. Therefore, if two individuals are both ``users'' or both ``non-users,'' we assume that they would develop a strong tie over time. 
Based on this rationale, we also model the interventionist who will exert a positive influence as an additional ``non-user'' who develops a weak tie to the ``users'' and a strong tie to the ``non-users,'' during the intervention period. 
An important component of the proposed intervention is to develop skills that would facilitate bonding with pro-social peers and discourage interactions with youth who have high-risk behaviors \cite{sussman1998one}. Therefore, if two individuals are separated and at least one of them has ``user'' behavior, their tie weakens or is cut.  Table \ref{table:net_in_out} summarizes how we expect strength of ties to change in response to graph partitioning. In this table, the rows indicate the behavior of the pair of nodes under consideration, and the columns show their pre-intervention tie. The changes are defined as a function of the ties pre-intervention ($\sets E$), the behavior of the end nodes ($b_i$), and whether they are assigned to the same group or not.\\

\begin{example}
Consider the example network in Figure~\ref{fig:network}. In this network, black circles indicate "user" individuals and white circles  represent "non-user" people. Initially, individuals are connected with either weak or strong ties, where weak ties are shown via light arrows, and strong ties are depicted via bold arrows. 

Given this social network as input, we need to group the individuals into two intervention groups. Based on this decision, new connections will form. Consider the grouping as depicted in Figure~\ref{fig:network}. Post-intervention, the ties undergo changes according to table~\ref{table:net_in_out}. Therefore, new ties will be formed between individuals that did not have any ties before intervention and are placed in the same group. Also, some of the weak existing ties become stronger before the intervention. Also, some of the weak ties will become stronger as a result of the grouping. Finally, the existing ties that are across groups will be broken or weakened.
\end{example}

\textbf{Substance Abuse Prevention Influence Spread Model. }
\label{par:Com_Inf_model}
 Following the intervention, and depending on how the network evolves, we evaluate the influence to predict the changes in the nodes' behaviors. We use a variant of the popular Linear Threshold model \cite{kempe2003maximizing} proposed in \cite{borodin2010threshold}, in which they study the competitive influence processes in social networks. Our problem is similar to \cite{borodin2010threshold} as there are two concurrent influences, a positive influence which originates from the ``non-users'' and a negative influence from ``users.'' However, it differs from their model in that the nodes in our network have already adopted one of these two behaviors.

Consider the post-intervention graph as $\sets G^+(\sets P) = (\sets V,\sets E^+(\sets P))$. 
For notational convenience, we henceforth eliminate the dependence of $\sets G^+$ and $\sets E^+$ on $\sets P$. 
For each $(i, j) \in \sets V \times \sets V$, a weight value $w_{ij} \in \mathbb R$ is assigned, which characterizes the amount of influence from node~$i$ to node~$j$. These weights are such that the sum of the weights of the incoming edges to a certain node is bounded. Specifically,
$$
0 \leq \sum_{i \in \sets V} w_{ij} \leq 1, \quad \forall j \in \sets V,
$$
%
Initially, each node $i \in \sets V$ is assigned a threshold value, denoted by ${\random T^i}$, corresponding to the influence threshold for switching behavior (from ``user'' to ``non-user'' and vice-versa). We assume these thresholds are uniformly distributed in the range $[0,1]$. A necessary condition for a node to change his behavior is for the sum of his neighbors' weights with opposite behavior to exceed the node's threshold. Mathematically, if for example, $b_j = 1$ and $\sum_{i : b_i=0}{w_{(i, j)}} \geq \random T^{j}$, node~$j$ has the necessary condition to become a ``non-user.'' Following the model from~\cite{borodin2010threshold}, if the threshold of a ``user'' (resp.\ ``non-user'') is exceeded, the node behavior will switch with probability $\Omega_{\text{u}|\text{n}}$ (resp.\ $\Omega_{\text{n}|\text{u}}$). Thus, for any node $i$, we can express their post-intervention behavior distribution as:
$$
P(\random B_j = 1) = \Omega_{\text{u}|\text{n}}  P\left(\sum_{i : b_{i} = 0}w_{i j} \geq \random T^{j}\right)\text{,  if }(b_j = 0)$$
$$ P(\random B_j = 0) = \Omega_{\text{n}|\text{u}} P\left(\sum_{i : b_{i} = 1}w_{ij} \geq \random T^{j}\right)\text{,  if }(b_j = 1).
$$





\begin{theorem}\label{thm:NP_hard}
Problem~\eqref{eq:general_formulation} is NP-Hard, even under the presented influence model. \end{theorem}

\begin{proof}
We construct a reduction from Set Cover problem with $N$ items to be covered
by $M$ subsets. The reduction can be explained as follows: 

For any item in the set cover
problem, a "non-user" node, and for any set, a "user" node will be created. There is a tie from a "user" to a "non-user", if and only if the item is included in a set. In the
drug prevention problem, we aim to partition the network into S sub-graphs. However,
several assumptions are made. First, all ties across partitions are cut. Every pair of "user" in the same group form a tie. Every member of the network is connected to a
"non-user" (the interventionist). All the ties have the same weight. The threshold values
for "non-users" are set to $\frac{\text{Number of "user" connections}}{
\text{Number of "user" connections + 1}}$. "User" thresholds are set equal to $\frac{1}{2}$ + $\epsilon$ ($\epsilon$ > 0 , small positive number.) The capacity of each group is N + M.
If the optimal solution is at least $(N + S + 1)$, there is at most one group with more
than two "users". This is because, if there are at least $N +S+ 1$ "non-user" at the end
of the intervention, at least S "users" must have flipped to "non-user", and the way
the thresholds are set requires that the "users" are not put together in a group, in more
that one group. Also, it is best to separate the "users" from "non-user". Therefore, the
optimal S "user" to isolate will be equivalent to the solution of the Set Cover. Also, a
solution to the Set Cover results in at least $N + S + 1$ "non-user".
\end{proof}

\section{Mixed-Integer Programming Formulation}\label{sec:integer_prog}

We present a MILP formulation of Problem~\eqref{eq:general_formulation}. First, we define $s^{\rm s}_{ij}, s^{\rm w}_{ij}, s^{\emptyset}_{ij}$ as indicators for whether the pre-intervention tie from $i$ to $j$ is strong, weak, or there is no tie, respectively. The values for these variables are known from the initial data on the network structure and are determined as $s_{ij}^\emptyset := \mathcal I ( (i,j) \notin \sets E  )$, $ s_{ij}^{\rm s} := \mathcal I ( (i,j) \in \sets E \text{ and }  s_{(i,j)} = 1)$, and $ s_{ij}^{\rm w} := \mathcal I ( (i,j) \in \sets E \text{ and }  s_{(i,j)} = 0)$, for each $(i,j) \in \sets V \times \sets V$, where $\mathcal I(\cdot)$ is the indicator function. 

Let us first introduce the decision variables. We define $z_{ij}$ as a binary variable which is equal to 1 if nodes $i$ and $j$ are in the same group. Also, let $x^{\rm s+}_{ij}, x^{\rm w+}_{ij}, x^{\emptyset+}_{ij}$ be binary variables to encode the post-intervention tie strength. Finally, $w_{ij}$ is the normalized post-intervention weight of the tie between~$i$ and~$j$. Next, we introduce the constraints in this problem.

The groups must satisfy the capacity constraints, that is the sum of all the nodes that are in the same group must meet the following constraint,i.e.,
$$
\underline C \; \leq \; \sum_{i \in \sets V}{z_{ij}} \; \leq \; \overline C \quad \forall j \in \sets V.
$$


we require that if i and j are in the same group, and so are j and k, i and
k must be in the same group as well. This can be expressed by saying that: 
$$
 z_{ij} + z_{jk} - 1 \; \leq \; z_{ki} \quad \forall i, j, k \in \sets V.
$$

The post-intervention link strengths can be defined dependent on the pre-intervention edge strengths, the group assignments, and the behavior of the nodes. Therefore, according to Table~\ref{table:net_in_out}, the link from $i$ to $j$ will become strong if for example, nodes $i$ and $j$ belong to the same group and either have similar behaviors, or they already have a strong tie, which is encoded as part of the following equation 
$$
\renewcommand{\arraystretch}{1.5}
\begin{array}{ccl}
{x^{\rm s+}_{ij}} & = & z_{ij} \left[ s^{\rm s}_{ij} + {|b_{i} + b_{j} - 1|}  (s^{\emptyset}_{ij}+s^{\rm w}_{ij})\right] \\ 
& & +(1-z_{ij}) {|b_{i} + b_{j} - 1| s^{\rm{s}}_{ij}},
\end{array}
$$
where the term $|b_{i} + b_{j} - 1|$ is 1 iff $i$ and $j$ have the same behavior. Similarly,
$$
\renewcommand{\arraystretch}{1.5}
\begin{array}{ccl}
{x^{\rm w+}_{ij}} & = & z_{ij} \left[ |b_{i} - b_{j}| (s^{\emptyset}_{ij}+s^{\rm w}_{ij})\right] + (1 - z_{ij}) \\ & &\left[s^{\rm w}_{ij} ({1 - \max(b_{i}, b_{j})}) + (s^{s}_{ij}) (|b_{i} - b_{j}|)\right].
\end{array}
$$
determines whether the new tie from $i$ to $j$ is weak.
Once the post-interventions network is determined, the normalized weights are calculated according to equation:
\renewcommand{\arraystretch}{1.5}
$$
w_{ij} =
\frac{W_{s} x^{\rm s+}_{ij} + W_{w} x^{\rm w+}_{ij}}{\sum_{i' \in \sets V} W_{s}  x^{\rm s+}_{i'j} + W_{w} x^{\rm w+}_{i'j}},
$$
where $W_{s}$ and $W_{w}, (W_{w} < W_{s})$ are numerical values for strong and weak ties, respectively. In this equation, the numerator is the strength of the tie from $i$ to $j$, before normalization which is equal to $W_{s}$ if it is a strong tie, or $W_{w}$ if it is a weak tie. The denominator is the sum of all the incoming weights.
While the above constraint consists of nonlinear terms, they can be linearized using standard techniques which require introducing a new decision variable for each product term. 

Finally, the objective can be defined as follows:
\begin{flalign*}
\label{eq:MIP_objective}
\sum_{j : b_{j} = 0}{\bigl(1-(\Omega_{\text{u}|\text{n}}\sum_{i: b_{i} = 1}{w_{ij}})\bigr)}+\sum_{j: b_{j}=1}{(\Omega_{\text{n}|\text{u}}\sum_{i: b_{i} = 0}
{w_{ij}})}.
\end{flalign*}
This equation gives the closed form objective to our problem, which is the expected number of ``non-users'' post-intervention. The first term in the above equation is the expected number of ``non-users'' that remain ``non-users.'' The second term is the expected number of ``users'' that change. 
This equation is equivalent to the objective in Problem \eqref{eq:general_formulation}. Note that since in the influence model, the thresholds are drawn from a uniform distribution, the probability that a node changes its behavior equals the normalized incoming edge weights from neighbors with that behavior. 
\section{Large Neighborhood Search (LNS)}
Local search algorithms target a set of feasible candidate solutions (neighborhood) and iteratively improve a solution by searching within that limited space. 
Small neighborhoods are faster to explore, but can make escaping a local optimum much harder. Large neighborhood search which was first proposed by \cite{shaw1998using}, can alleviate this problem by exploring a carefully chosen large neighborhood. LNS iteratively improves a solution by alternatively destroying and repairing a solution.

Our solution approach is based on a mixture of LNS and MILP optimization which has been proven successful in \cite{lim2015large}. Starting from an initial feasible solution, each loop starts by destroying the current solution by selecting two groups at random. It then repairs the solution by optimally deciding how to partition people between these two groups. Since this MILP subproblem is much smaller than the original MILP, it yields a significant improvement in the run time. 
It is important to note that in the destroy step, determining the amount of destruction is crucial. If too little is destroyed, the effect of large neighborhood is lost, whereas large destructions will result in a repeated re-optimization. We experimented with different neighborhoods, for example destroying the group assignment of random nodes or three groups at the same time (instead of two). In these cases, the results were not performing as well as the two-group neighborhood.


\section{Experimental Setup}
\textbf{Network Data. } Social Network data primarily comes from the Youthnet study in which homeless youth were recruited from a drop-in center in Los Angeles County, CA. 
Clearly, our approach is independent of the input network data. Therefore, we use this dataset for our preliminary study. We also plan to deploy our approach in Denver and in collaboration with Urban Peak organization. 

In order to explore different network structures, we use randomly generated Watts Strogatz (WS) graphs \cite{watts1998collective} with parameters ($p = 0.25$, $k = 4$). WS graph models have properties such as short average path lengths which makes them close to real-world networks. \\
\textbf{Baselines. } For evaluation, we compare MILP and LNS against three different baselines. These baselines are based on the three approaches that were detailed in the Introduction section (random assignment, network-based or by participants choice, and teacher nominated assignments).\\
\textbf{Model Parameters. } In the following experiments, and based on input from domain experts we choose the model parameters as: 
$\Omega_{\text{u}|\text{n}} = 1.0$, $\Omega_{\text{n}|\text{u}} = 0.8$, $\underline C = 3$, $\overline C = 8$, $W_{s} = 3$, and $W_{w} = 1$. We also explore the impact of variations of some of these parameters.
For the LNS, 50 trials are performed, each starting from a different random initial solution and we report the solution with highest expected \textit{success} rate. We also report averages over 25 different graph samples, both for the real and WS graphs. 

\section{Results and Discussion}
\textbf{Solution Quality Metrics. } 
Different solution strategies are compared based on a \textit{success} metric, which we define as:
$$
\textit{success} = \frac{E(\sum_{i \in \sets V}{(1-\random B_i(\sets P))}) - \sum_{i \in \sets V}{(1 - b_{i})}}{\max \bigl( E(\sum_{i \in \sets V}{(1-\random B_i(\sets P))}) - \sum_{i \in \sets V}{(1 - b_{i})} \bigr)}
$$
The numerator is the expected number of youth that have become ``non-users'' as the result of the intervention. The term~$E(\sum_{i \in \sets V}{(1-\random B_i(\sets P))})$ corresponds to the objective function of Problem~\eqref{eq:general_formulation}. The denominator is its maximum possible value which is bounded above by $\Omega_{\text{n}|\text{u}} \sum_{i \in \sets V}b_{i}$ -- it corresponds to the case where all ``users'' thresholds are exceeded, and thus the maximum expected number of ``users'' that will change.\\
\textbf{Neighborhood Search: Small vs. Large. }
We first compare the \textit{success} rate of the two local search based algorithms. In each step of the small neighborhood search, a randomly chosen pair of nodes switch groups. If the resulting solution has a higher \text{success} rate, the solution is accepted and this process continues until a locally optimal solution is found. 
\begin{figure}
\includegraphics[width=5cm]{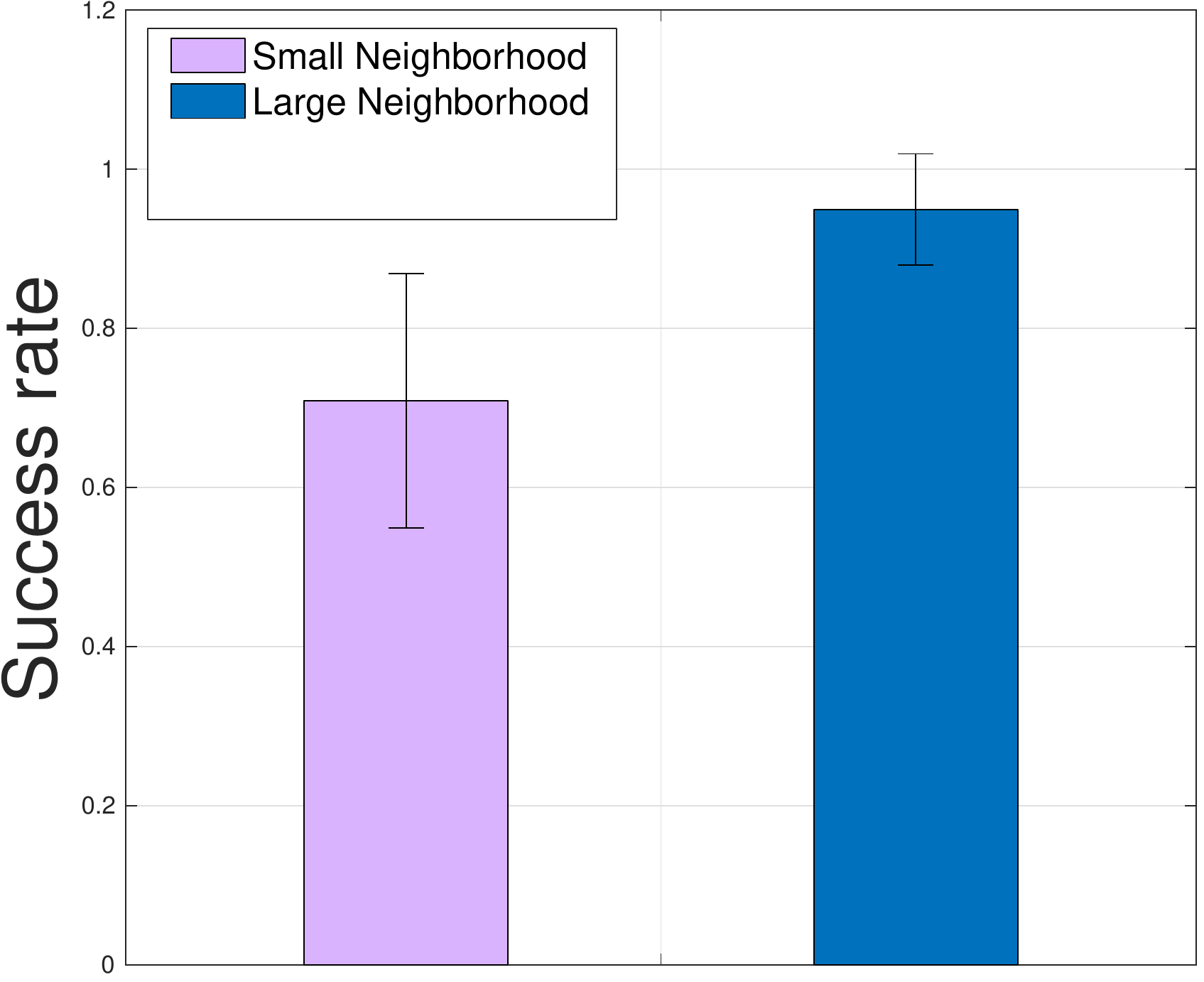}
\caption{{Comparison between small and large neighborhood search}}
\label{fig: small_large}
\end{figure}
As explained previously, the LNS implementation is based on re-optimization of the randomly chosen groups. 
For both algorithms, each run is repeated for different initial random solutions until a time limit is reached. In this experiment, we use 10 different randomly  generated WS graphs of size 30. The time limit is set to 1 hour. 
Figure \ref{fig: small_large} shows that the small  neighborhood search can perform poorly relative to LNS under a variety of instances. This results indicates that between MILP, which is not easily scalable to larger networks, and pure small neighborhood local search which is favorable due its lower computational demand, LNS can achieve a balance between quality and speed and it is able to outperform local search, given the same time budget. 

\begin{figure} \label{fig: sol_Quality}
  \centering
  \includegraphics[width=5cm]{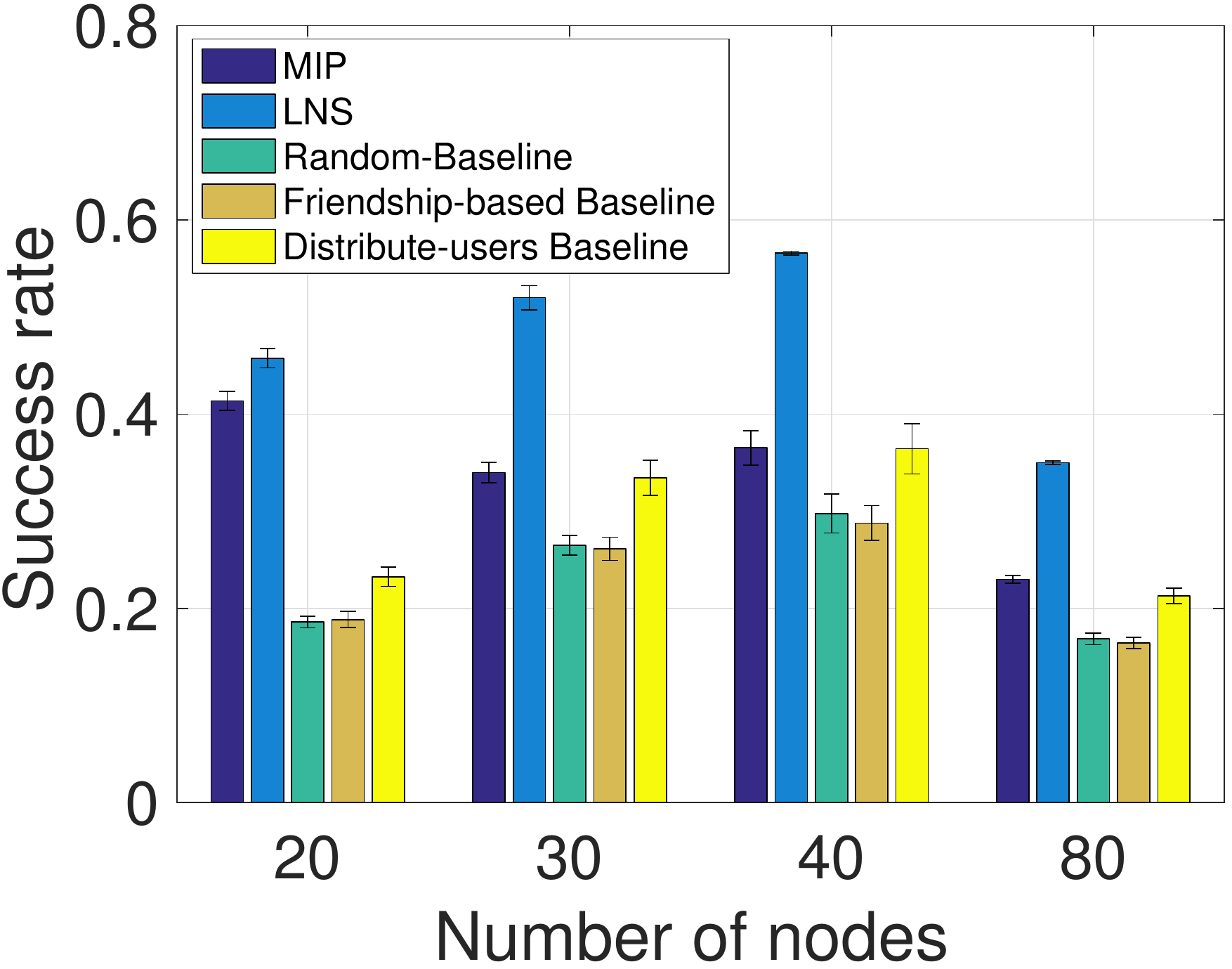}
 \caption{Solution quality of MIP and LNS in GUIDE, against three baselines commonly employed by practitioners (Synthetic Graphs)}
  \label{fig: sol_Quality_syn}
\end{figure}

\begin{figure}
  \includegraphics[width=5cm]{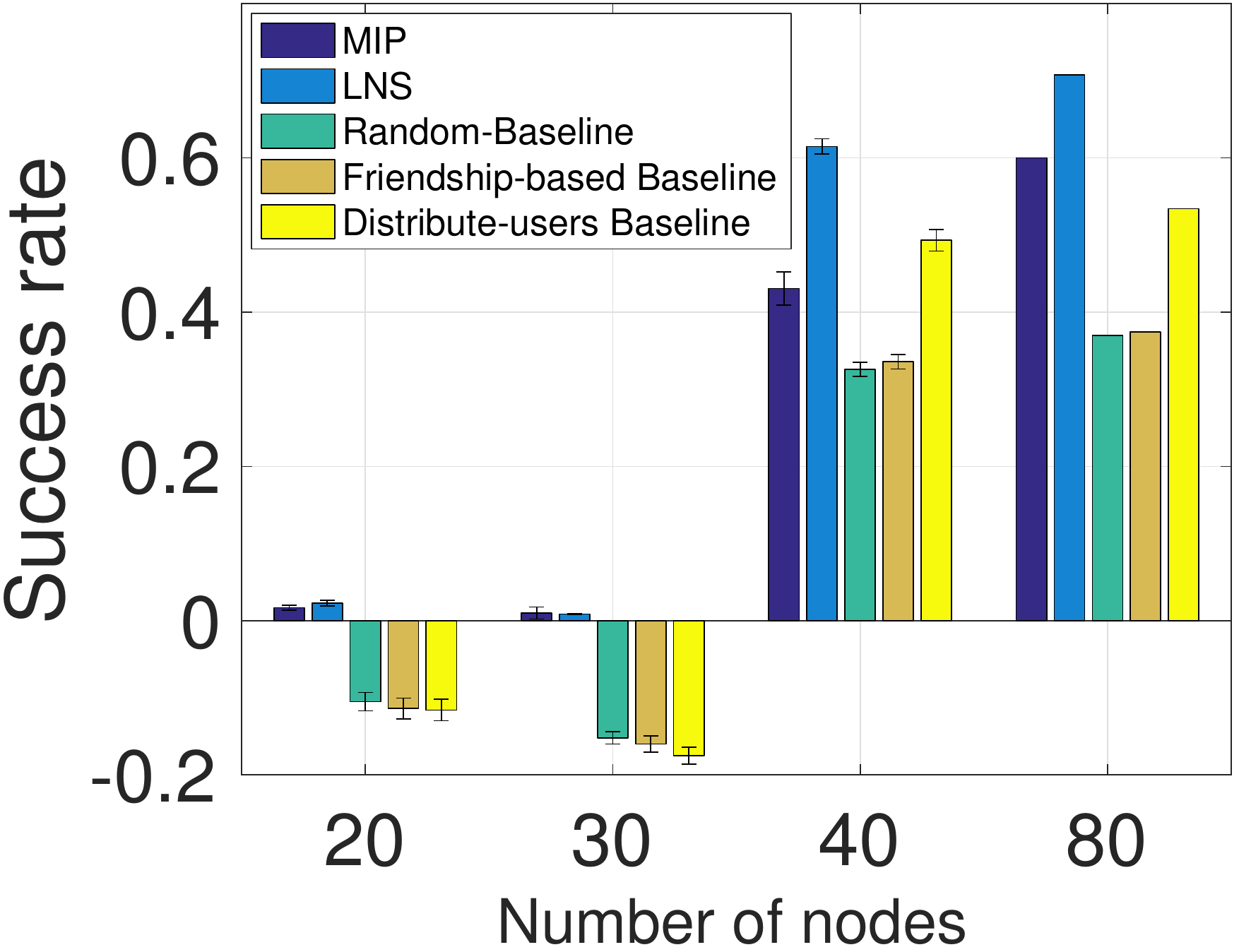} \\
  \caption{Solution quality of MIP and LNS in GUIDE, against three baselines commonly employed by practitioners (Real Graphs)} 
  \label{fig: sol_Quality_real}
\end{figure}

\textbf{Solution Quality. } Figures~\ref{fig: sol_Quality_syn} and~\ref{fig: sol_Quality_real} compare the \textit{success} rate of the optimization techniques, MIP and LNS against our baselines across~4 different network sizes, for both real network samples and WS samples. The MIP solver is given a cutoff time equal to the solution time of LNS (summed over the 50 trials).  
These results indicate that the solutions of both MIP and LNS are significantly better than any of the traditional methods for forming these groups, both statistically $(p < 0.01)$ and practically. From the practical point of view, prevention science cares about effect size of interventions in addition to the statistical significance, and we observed in the results that we were able to provide solutions that outperform the traditional methods up to (20 - 41)\%.
Figure \ref{fig: sol_Quality_real} also shows that in the real network samples (sizes 20 and 30), the baselines have a negative \textit{success} rate. Negative \textit{success} rate corresponds to the deviancy training effect, which means the expected number of ``non-users'' post-intervention is fewer than pre-intervention. This observation can be explained by the structure of the network as well as the behavior of the nodes. In fact, in these samples, a significant ratio of the individuals are ``users'' (68\%), hence the negative influence dominates. Additionally, many of these ``users'' are strongly connected with their ``user'' peers, making it hard for them to be positively influenced. We should note that the optimized strategies are always strictly positive, though the \textit{success} rate is limited due to the certain network structure. 
It can also be observed that the common intuition of evenly distributing the ``users'' across groups is in fact performing relatively well as compared to the other baselines. The importance of this result is twofold; First, it validates our model by reflecting the intuition of the social work partners that such a strategy is helpful. Secondly, we can clearly see that there is room for further improvements over these baselines. 


\begin{figure}
 \includegraphics[width=5cm]{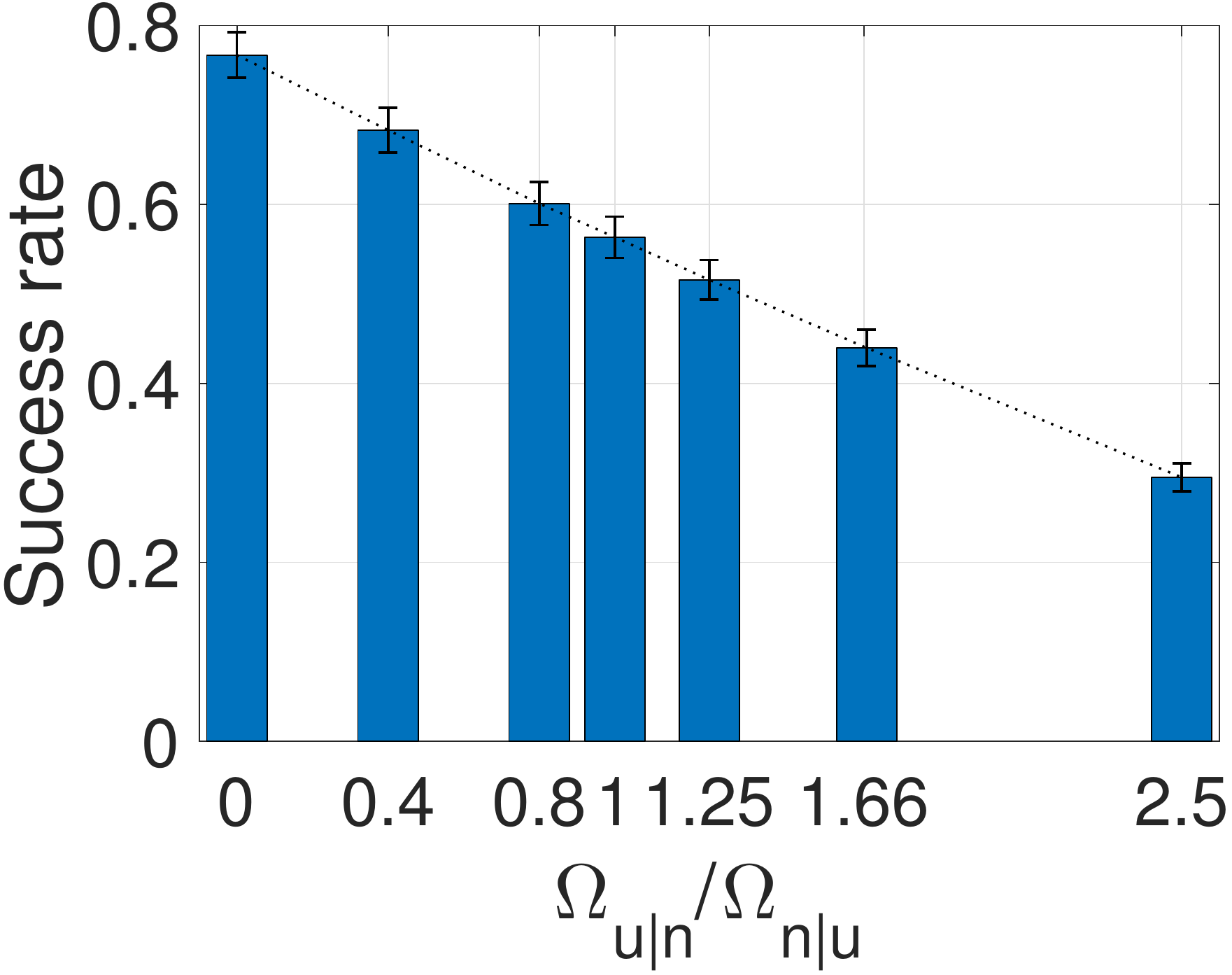}
 \caption{\textit{Success} vs. ${\Omega_{\text{u}|\text{n}}}/{\Omega_{\text{n}|\text{u}}}$}
 \label{fig: ratio}
\end{figure}

There are computational lessons learned as well. 
For example, MIP is guaranteed to find the optimal solution by searching the entire solution space, but as shown here, it is not a practical solution due to time constraints. 
In fact, LNS outperforms MIP solution given the same time budget. 
Figure~\ref{fig: ratio} compares the intervention \textit{success} with respect to different $\frac{\Omega_{\text{u}|\text{n}}}{\Omega_{\text{n}|\text{u}}}$ ratios.
This experiment is performed on synthetic graphs of size 40.
It illustrates that the increase of this ratio will result in a decline in expected \textit{success} of our intervention. 
\\

\textbf{Scalability. } 
\begin{figure}
  \centering
  \includegraphics[width=5cm]{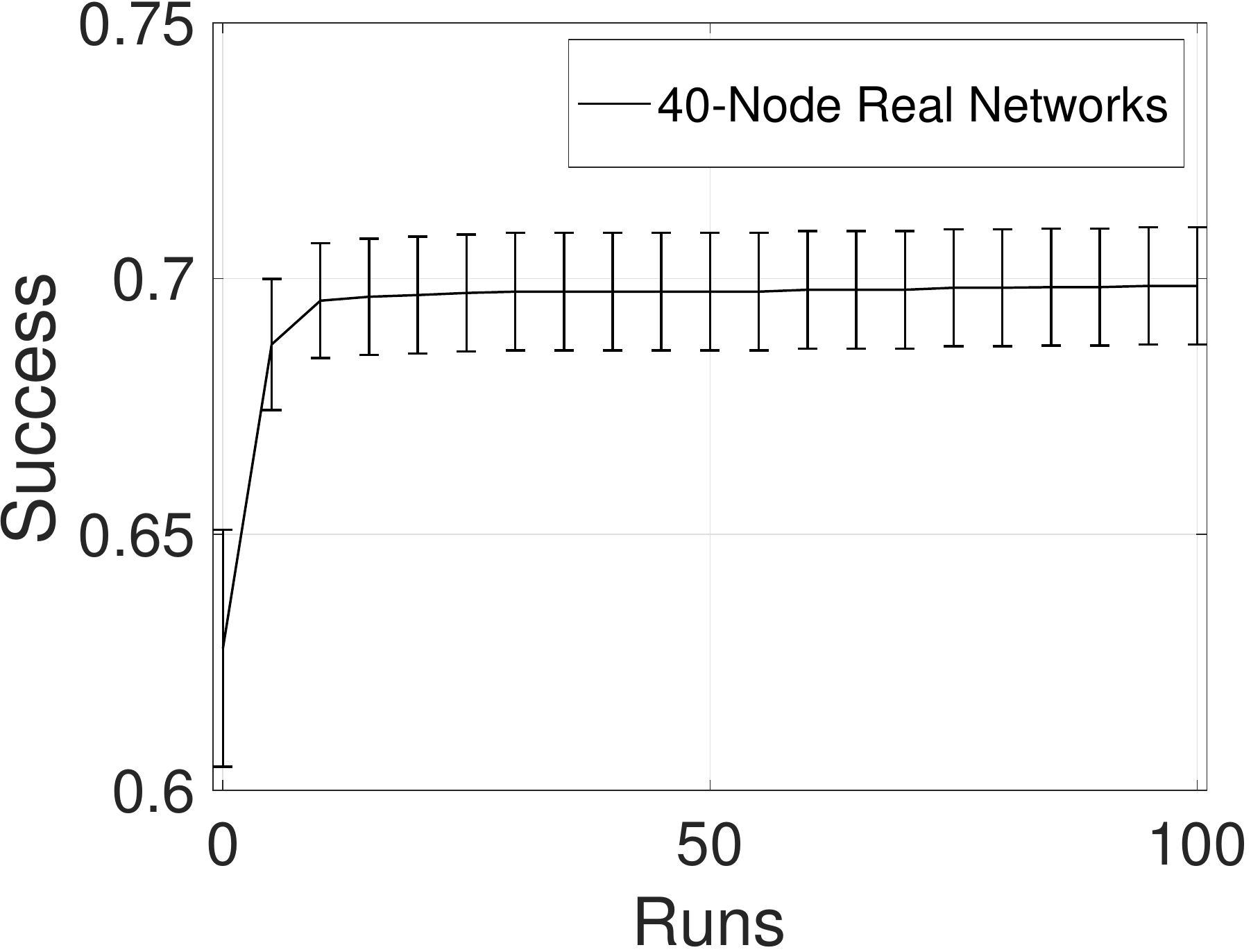}
\caption{LNS solution time analysis. This plot indicates how the solution evolves with respect to the number of runs.}
\label{fig: MIPscale_Hillscale}
\end{figure}

\begin{figure}
\includegraphics[width=5cm]{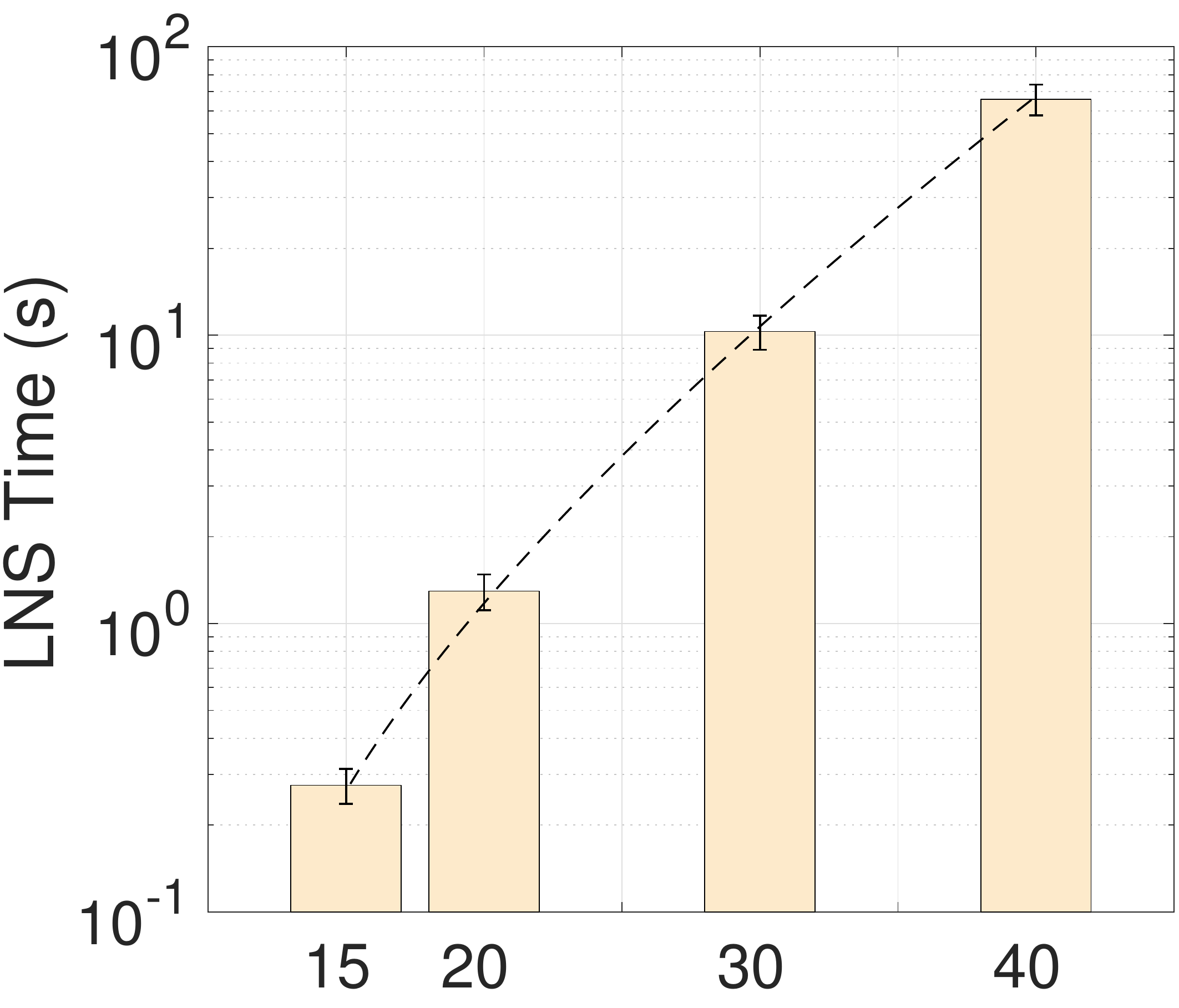}  
\caption{LNS solution time analysis. This plot shows how the time required to find a locally optimal solution changes with respect to the network size.}
  \label{fig: MIPscale_Hillscale_1}
\end{figure}

Figure \ref{fig: MIPscale_Hillscale} shows the LNS solution quality in a 40-node real-world graphs against the number of runs. Recall that LNS is run for 50 times, each starting from a different random initial solution. Figure \ref{fig: MIPscale_Hillscale} (a) shows that almost after 20 runs, the solution does not change significantly.
Figure \ref{fig: MIPscale_Hillscale_1} shows the time needed for the LNS to find a locally optimal solution for different network sizes. This plot shows that as the size of the network increases, the LNS time increases polynomially. These results suggest that LNS is an efficient choice for our purpose, mainly because a decision aid, such as GUIDE, needs to be adjustable to unanticipated occurrences. For example in cases a person does not show up, or refuses to accept his assignment, GUIDE needs to recalculate a new solution relatively quickly based on the imposed constraints. Therefore we will utilize LNS as the core optimization algorithm in GUIDE.

\section{Deployment}
GUIDE is developed in collaboration with Urban Peak,
a homeless-youth serving organization in Denver, CO, and it is under preparation for deployment. We have developed an application that encapsulates the data collection phase as well as the implementation of the LNS, the optimization algorithm. This application has a user-friendly interface that facilitates the use of our algorithm for the practitioners. 

\begin{figure}
  \centering
  \begin{tabular}{c c}
    \includegraphics[width=.45\linewidth]
{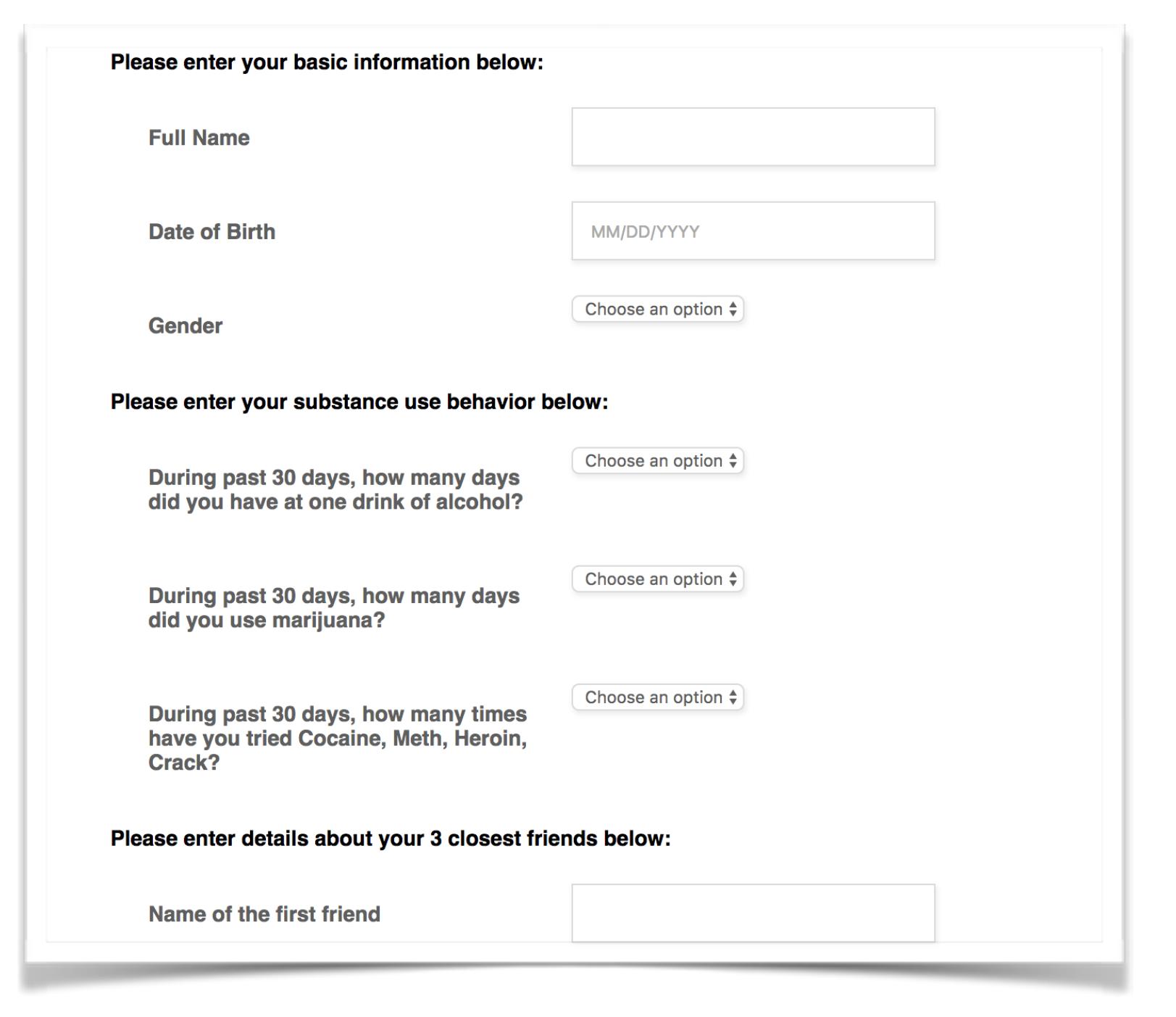} & 
    \includegraphics[width=.45\linewidth]{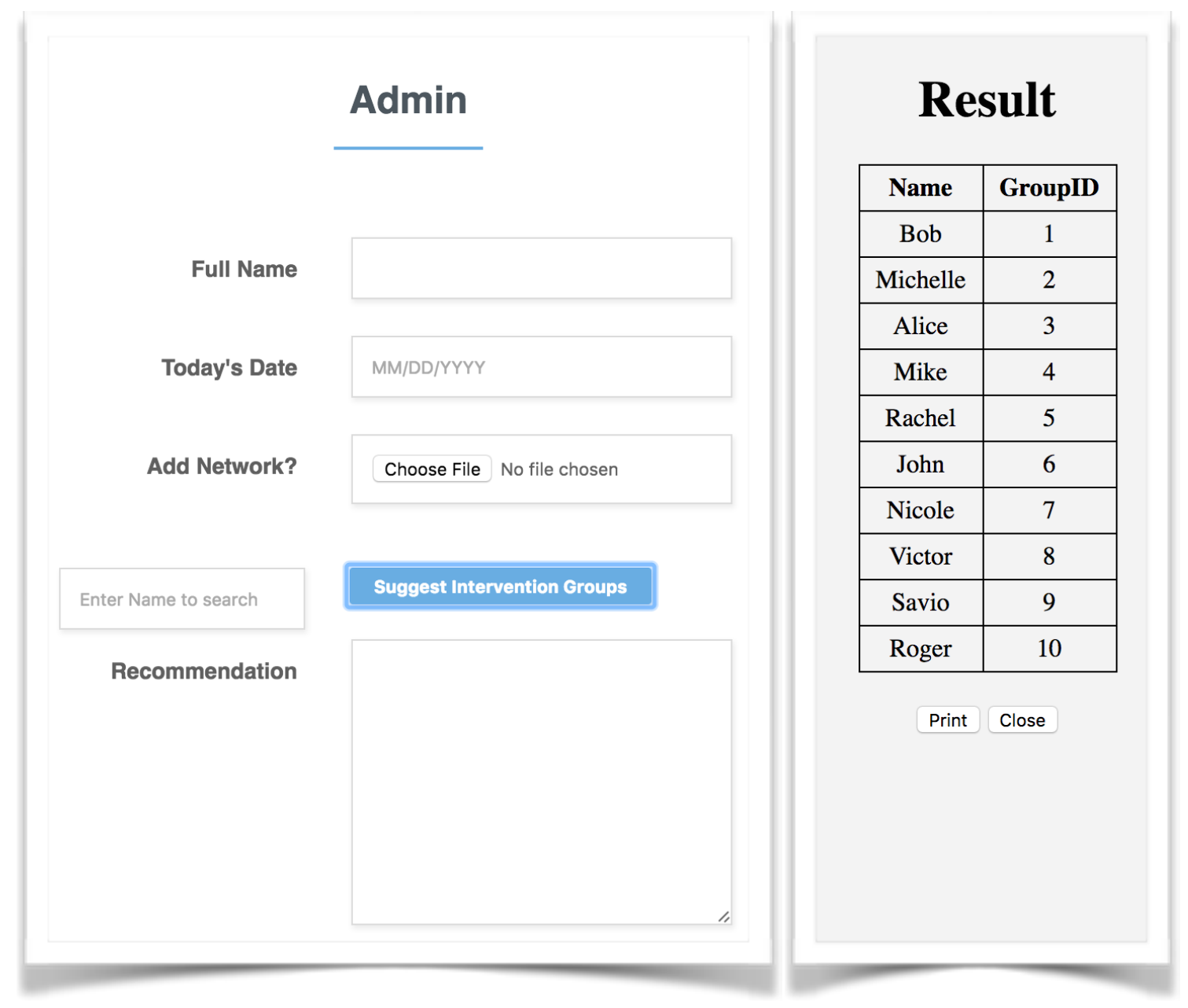}  \\ \small (a) A snapshot of the & \small (b) A snapshot of the GUI 
 \\ \small data collection form  & \small for the group recommendation \\
  \end{tabular} \caption{Two snapshots of the GUIDE application. (a) First, detailed information about the substance-use behavior of each individual is collected and saved in a data base. (b) Second, the practitioners will query the application for a group recommendation.}\label{fig: application}
\end{figure}

In the first phase, detailed data about the social network and substance use behavior of each volunteer is collected. Then, a personalized ID is assigned to each individual and the data is stored on the local machine for future use. See Figure \ref{fig: application} (a). 
Next, the data is fed into the LNS algorithm to generate a suggestion for intervention groups. The practitioners can easily upload or choose an existing social network data file. After the optimization is performed, the results will appear on the screen, See Figure \ref{fig: application} (b). 
The results include a group assignment for each individual which is formatted as a table. Finally, the output can be printed and saved to a file for future references. 

\section{Conclusion}
Substance abuse is a very significant public health and social problem in the United States. We have identified a niche in this wicked social problem where AI can make a contribution, illustrated the many modeling complexities, and provided algorithmic improvements over current practice. The result is GUIDE, an AI-based decision aid which leverages social network to structure intervention groups. We evaluated our approach against different traditional methods ranging from those based on intuitions such as distributing ``users'' to purely random strategies with no effort to understand the social circle of these youth. We showed that these traditional strategies significantly under-perform relative to our proposed method, emphasizing the importance of social
influence in composition of these groups. In fact, here our focus is on the life choices of a extremely vulnerable section of the population. The result, as we discussed, is a model that is of tremendous value to prevention scientists, providing them with a quantitative means of understanding phenomena for which until now only qualitative observations were feasible. In particular, as these interventions
are very costly and time consuming, the ability to forecast the likely impact of different strategies is very helpful.


\bibliographystyle{ACM-Reference-Format}
\bibliography{paper}

\end{document}